\documentclass[a4paper]{jpconf}
\usepackage{graphicx}
\usepackage{amsmath,amssymb}

\begin{document}

\title{Superradiance and quantum states on black hole space-times}

\author{Visakan Balakumar$^1$, Rafael Bernar$^2$ and Elizabeth Winstanley$^1$}

\address{$^1$ Consortium for Fundamental Physics,
	School of Mathematics and Statistics,
	The University of Sheffield,
	Hicks Building,
	Hounsfield Road,
	Sheffield. S3 7RH United Kingdom}

\address{$^2$ Faculdade de F\'isica, 
Universidade Federal do Par\'a, 
66075-110, Bel\'em, Par\'a, Brazil} 

\ead{V.Balakumar@sheffield.ac.uk}
\ead{rbernar@ufpa.br}
\ead{E.Winstanley@sheffield.ac.uk}

\begin{abstract}
We consider the definition of the Boulware and Hartle-Hawking states for quantum fields on black hole space-times. The properties of these states on a Schwarzschild black hole have been understood for many years, but neither of these states has a direct analogue on a Kerr black hole. We show how superradiant modes play an important role in the definition of quantum states on Kerr.  Superradiance is also present on static black hole space-times, in particular for a charged scalar field on a Reissner-Nordstr\"om black hole. We explore whether analogues of the Boulware and Hartle-Hawking states exist in this situation. 
\end{abstract}

\section{Quantum field theory on curved space-time}
\label{sec:intro}

Quantum field theory on curved space-time is a semiclassical approach to quantum gravity, wherein the space-time is fixed and purely classical. 
The matter content in the theory consists of quantum fields, which propagate on the classical background geometry.
One of the most famous results in this framework is the discovery by Hawking \cite{Hawking:1974sw} that black holes formed by gravitational collapse emit thermal quantum radiation with temperature $T_{H}$ given, in natural units by
\begin{equation}
T_{H} = \frac{\kappa }{2\pi } ,
\label{eq:temp}
\end{equation}
where $\kappa $ is the surface gravity of the black hole.

The simplest type of black hole is described by the Schwarzschild metric
\begin{equation}
ds^{2} = -\left( 1 - \frac {2M}{r} \right) dt^{2}
+ \left( 1- \frac {2M}{r} \right) ^{-1} dr^{2}
+ r^{2} \, d\theta ^{2} + r^{2} \sin ^{2} \theta \, d\varphi ^{2} ,
\label{eq:Schwarzschild}
\end{equation}
where $M$ is the mass of the black hole (we employ units in which $G=c=\hbar=k_{{\mathrm{B}}}=1$). On this background geometry, three standard quantum states have been constructed (see Section~\ref{sec:Schwarzschild} for more details).
The first is the Unruh state \cite{Unruh:1976db}, which contains, at future null infinity, an outgoing flux of Hawking radiation.
 A key feature of the Unruh state is that it is regular across the future horizon of the black hole.
In contrast, the Boulware state \cite{Boulware:1974dm} contains no particles as seen by a static observer far from the black hole.
Finally, the Hartle-Hawking state \cite{Hartle:1976tp,Israel:1976ur} represents a black hole surrounded by thermal radiation at the Hawking temperature.
This state is regular everywhere on and outside the event horizon. 
Our purpose in this work is to investigate the analogues of these states for quantum scalar fields on Reissner-Nordstr\"om and Kerr black holes. 
We begin by briefly outlining the definition of quantum states for a neutral scalar field via canonical quantization, and how this process is applied on a Schwarzschild black hole.
In Section~\ref{sec:Kerr} we consider a neutral scalar field on a Kerr black hole space-time, and in Section~\ref{sec:RN} our focus is the Reissner-Nordstr\"om black hole, but with a charged rather than neutral scalar field.
Our conclusions are presented in Section~\ref{sec:conc}.

\subsection{Canonical quantization of a neutral scalar field}
\label{sec:neutral}

Let $\Phi $ be a neutral scalar field. 
The process of canonical quantization begins by expanding the classical field in an orthonormal basis of field modes, which we write schematically as follows:
\begin{equation}
\Phi = \sum _{j} a_{j}\phi _{j}^{+} + a_{j}^{\dagger }{\phi _{j}^{-}} ,
\end{equation} 
where $j$ represents the quantum numbers characterizing the field modes.
Here the $\phi _{j}^{\pm }$ modes are, respectively, positive and negative frequency modes.
For example, in two-dimensional Minkowski space-time where the modes are given by plane waves, in terms of time $t$ and a spatial coordinate $x$, the modes $\phi _{j}^{\pm }$ take the form
\begin{align}
  \phi _{j}^{+} & \propto e^{-i\omega (t\pm x)},  \qquad \omega >0 ,
  \nonumber \\
  \phi _{j}^{-} & \propto e^{-i\omega (t \pm x)},  \qquad \omega <0,
\end{align}
where $\omega $ is the frequency of the mode.
The field is quantized by promoting the expansion coefficients $a_{j}$, $a_{j}^{\dagger }$ to operators satisfying the commutation relations
\begin{equation}
\left[ {\hat {a}}_{j}, {\hat {a}}_{k}^{\dagger } \right] = \delta _{jk},
\qquad
\left[ {\hat {a}}_{j}, {\hat {a}}_{k} \right] = 0,
\qquad
\left[ {\hat {a}}_{j}^{\dagger }, {\hat {a}}_{k}^{\dagger } \right] = 0.
\end{equation}
The vacuum state $\left| 0 \right\rangle$ is then defined as that state annihilated by the ${\hat {a}}_{j}$ operators, that is: ${\hat {a}}_{j} \left| 0 \right\rangle =0$ for all $j$. 
Therefore the definition of positive and negative frequency modes plays a central role in the construction of quantum states.

\subsection{Schwarzschild black hole}
\label{sec:Schwarzschild}

We now briefly outline the canonical quantization of a neutral scalar field on a Schwarzschild black hole with metric (\ref{eq:Schwarzschild}). 
Mode solutions of the neutral scalar field equation on this background take the form
\begin{equation}
\phi _{\omega \ell m }(t,r,\theta, \varphi ) =
\frac {1}{{\mathcal {N}}r} \,
e^{-i\omega t} \, e^{im \varphi } \, Y_{\ell m}(\theta )
R_{\omega \ell }(r) ,
\label{eq:Smodes}
\end{equation}
where ${\mathcal {N}}$ is a normalization constant, $\omega $ is the frequency of the modes with respect to Schwarzschild time $t$, the integer $m\in {\mathbb{Z}}$ is the azimuthal quantum number and $Y_{\ell m}(\theta )$ is a spherical harmonic with total angular momentum quantum number $\ell \in {\mathbb{Z}}^{+}$.  
The radial function $R_{\omega \ell }(r)$ satisfies a Schr\"odinger-like equation
\begin{equation}
\left[ \frac {d^{2}}{dr_{*}^{2}} + V_{\omega \ell}(r) \right] R_{\omega \ell }(r) = 0 ,
\label{eq:Sradial}
\end{equation}
where $r_{*}$ is the ``tortoise'' coordinate, defined by the differential equation
\begin{equation}
\frac {dr_{*}}{dr}  =   \left( 1-\frac {2M}{r} \right) ^{-1} .
\end{equation}
As $r\rightarrow r_{h}=2M$, the event horizon radius, the tortoise coordinate $r_{*}\rightarrow - \infty $, while $r_{*}\rightarrow \infty$ far from the black hole, as $r\rightarrow\infty$. 
In these asymptotic regions, the potential $V_{\omega \ell }(r)$ has the following behaviour:
\begin{equation}
V _{\omega \ell}(r) \rightarrow \omega ^{2} , \qquad r_{*}\rightarrow \pm \infty .
\end{equation}
Therefore, as $r_{*}\rightarrow \pm \infty $, we have mode solutions of the form $e^{-i\omega (t\pm r_{*})}$ with frequency $\omega $, and hence the natural definition of positive frequency with respect to Schwarzschild time $t$ is $\omega >0$. 
A suitable orthonormal basis of field modes is formed by the ``in'' and ``up'' modes, whose radial functions have the following forms in the asymptotic regions:
\begin{subequations}
\begin{align}
R_{\omega \ell }^{{\mathrm {in}}}(r)  & \sim \left\{
\begin{array}{ll}
B_{\omega \ell }^{{\mathrm {in}}} \, e^{-i\omega  r_{*}} 
& r_{*}\rightarrow -\infty ,
\\
e^{-i\omega r_{*}} + A_{\omega \ell }^{{\mathrm {in}}} \, e^{i\omega r_{*}}
& r_{*} \rightarrow \infty ,
\end{array}
\right.
\\
R_{\omega \ell }^{{\mathrm {up}}}(r)  & \sim 
\left\{
\begin{array}{ll}
e^{i \omega r_{*}} 
+ A_{\omega \ell }^{{\mathrm {up}}} \, e^{-i\omega  r_{*}} 
& r_{*}\rightarrow -\infty ,
\\
B_{\omega \ell }^{{\mathrm {up}}} \, e^{i\omega r_{*}}
& r_{*} \rightarrow \infty ,
\end{array}
\right.
\end{align}
\end{subequations}
where $A_{\omega \ell }^{{\mathrm {in/up}}}$ and $B_{\omega \ell }^{{\mathrm {in/up}}}$ are complex constants.
The ``in''  modes represent scalar waves incoming from past null infinity, which are partly reflected back to  future null infinity and partly transmitted down the future horizon.
The ``up'' modes represent scalar waves which are outgoing near the past horizon, partly reflected back down the future horizon and partly transmitted to  future null infinity.

Using the ``in'' and ``up'' modes, suitably normalized, and defining positive frequency with respect to Schwarzschild time $t$, the resulting vacuum state is the Boulware state $|{\mathrm {B}}\rangle $ \cite{Boulware:1974dm}.
However, Schwarzschild time $t$ is a good time coordinate only in the region outside the event horizon.
Alternatively, we can consider Kruskal time $T$, which is a valid time coordinate everywhere in the space-time. 
We can construct an orthonormal basis of modes having positive frequency with respect to Kruskal time $T$ from the ``in'' and ``up'' modes (see, for example, the construction in \cite{Novikov:1989sz}). 
The resulting vacuum state is then the Hartle-Hawking state $|{\mathrm {H}}\rangle $ \cite{Hartle:1976tp}.
In the definition of both the Boulware and Hartle-Hawking states, the ``in'' and ``up'' modes are defined to be positive frequency with respect to the same time coordinate.
In contrast, to construct the Unruh state $|{\mathrm {U}}\rangle $ \cite{Unruh:1976db}, the ``in'' modes are defined as having positive frequency with respect to Schwarzschild time $t$, while the ``up'' modes have positive frequency with respect to Kruskal time $T$. 

To understand the physical properties of these states, in this note we will consider unrenormalized expectation values of field operators, to avoid the technical challenges inherent in the renormalization process.
Since the divergent terms which need to be subtracted to give renormalized expectation values are purely geometric and independent of the quantum state (see, for example, \cite{Decanini:2005eg} for a neutral scalar field and \cite{Balakumar:2019djw} for a charged scalar field, which will be considered in Section~\ref{sec:RN} below), differences in expectation values between two quantum states do not require renormalization.
For a general operator ${\hat {O}}$ (for example, the square of the quantum scalar field or the scalar field stress-energy tensor ${\hat {T}}_{\mu \nu}$), the unrenormalized expectation values of the operator in the Boulware, Unruh and Hartle-Hawking states on Schwarzschild space-time can be written  as mode sums:
\begin{subequations}
\label{eq:SchwarzExp}
\begin{align}
\langle {\mathrm {B}} | {\hat {O}} | {\mathrm {B}} \rangle
& = 
\sum _{\ell = 0}^{\infty } \sum _{m=-\ell }^{\ell }
\int _{0}^{\infty } d{\omega}
\left[ o _{\omega  \ell m}^{{\mathrm {in}}} 
+ o _{\omega \ell m}^{{\mathrm {up}}}  \right] ,
\label{eq:BSchwarz}
\\
\langle {\mathrm {U}} | {\hat {O}} | {\mathrm {U}} \rangle 
& = 
\sum _{\ell = 0}^{\infty } \sum _{m=-\ell }^{\ell }
\int _{0}^{\infty } d{\omega}
\left[ o_{\omega  \ell m}^{{\mathrm {in}}} 
+ o _{\omega \ell m}^{{\mathrm {up}}}  \coth \left( \frac {{\omega }}{2T_{H}} \right) \right]  ,
\label{eq:USchwarz}
\\
\langle {\mathrm {H}} | {\hat {O}} | {\mathrm {H}} \rangle 
& = 
\sum _{\ell = 0}^{\infty } \sum _{m=-\ell }^{\ell }
\int _{0}^{\infty } d{\omega} 
\left[ o_{\omega  \ell m}^{{\mathrm {in}}} 
+ o _{\omega \ell m}^{{\mathrm {up}}} \right] 
\coth \left( \frac {{\omega }}{2T_{H}} \right) ,
\label{eq:HSchwarz}
\end{align}
\end{subequations}
where $T_{H}$ is the Hawking temperature of the black hole (\ref{eq:temp}), and $o^{\rm {in/up}}_{\omega \ell m}$ is the classical value of $O$ when calculated for a particular scalar field mode.
All the unrenormalized expectation values involve sums over the angular momentum quantum numbers $\ell $ and $m$, and integrals over positive frequency $\omega >0$.

We interpret the mode sums in (\ref{eq:SchwarzExp}) as follows.
The Boulware state $|{\mathrm {B}}\rangle $ contains no particles in either the ``in'' or the ``up'' modes, as seen by a static observer far from the black hole.
In contrast, the Hartle-Hawking state $|{\mathrm {H}}\rangle $ contains a thermal distribution of particles in both the ``in'' and the ``up'' modes, at temperature $T_{H}$. Note that the frequency $\omega $ of each mode appears in the thermal factor in the mode sum.
Finally, the Unruh state $|{\mathrm {U}}\rangle $ is devoid of ``in'' mode particles, but contains a thermal flux of particles in the ``up'' modes, corresponding to the outgoing Hawking radiation of the black hole.
 
\section{Neutral scalar field on Kerr black hole}
\label{sec:Kerr}

Having outlined the construction of the standard quantum states for a neutral scalar field on a static Schwarzschild black hole, we next consider a neutral scalar field on a rotating Kerr black hole, having metric
\begin{subequations}
    \label{eq:Kerr}
\begin{equation}
ds^{2} =
-\frac {\Delta }{\Sigma }
\!\!\left[ dt - a\sin ^{2} \theta \, d\varphi  \right]^{2}
+ \frac {\Sigma }{\Delta }dr^{2}
+ \Sigma \, d\theta ^{2}
+\frac {\sin ^{2} \theta }{\Sigma }
\!\!\left[ \left( r^{2} + a^{2} \right) d\varphi - a \, dt \right] ^{2} ,
\end{equation}
where the functions  $\Delta (r)$ and $\Sigma (r)$ are given by
\begin{equation}
		\Delta  =
		r^{2} -2Mr + a^{2} ,
		\qquad
		\Sigma  =
		r^{2} + a^{2} \cos ^{2} \theta ,
\end{equation}
\end{subequations}
and $M$ is the mass of the black hole, with $a$ the spin parameter.

\subsection{Neutral scalar field modes}
\label{sec:KerrModes}
The neutral scalar field equation on the background (\ref{eq:Kerr}) has mode solutions of the form
\begin{equation}
\phi _{\omega \ell m} (t,r,\theta, \varphi ) 
= \frac {1}{{\cal {N}}}\frac {1}{\left( r^{2}+a^{2}
\right) ^{\frac {1}{2}}}e^{-i\omega t} e^{im\varphi } S_{\omega \ell m}(\cos \theta )
R_{\omega \ell m} (r) ,
\end{equation}
where  ${\mathcal {N}}$ is a normalization constant, $S_{\omega \ell m}(\cos \theta )$ is a spheroidal harmonic and the angular quantum numbers $\ell $, $m$ are as in the Schwarzschild case. 
The radial function $R_{\omega \ell m}(r)$ again satisfies a Schr\"odinger-like equation
\begin{equation}
\left[ \frac {d^{2}}{dr_{*}^{2}} + V_{\omega \ell m}(r) \right]
				R_{\omega \ell m}(r) =0,
    \label{eq:radKerr}
\end{equation}
where the ``tortoise'' coordinate $r_{*}$ is now defined by 
\begin{equation}
\frac {dr_{*}}{dr}  =   \frac {r^{2}+a^{2}}{\Delta } .
\end{equation}
Close to the event horizon at $r=r_{h}$, $r_{*}\rightarrow -\infty $, and far from the black hole  ($r,r_{*}\rightarrow \infty $), the potential $V_{\omega \ell m}(r)$ takes the form
\begin{equation}
V_{\omega \ell m}(r)=\left\{
\begin{array}{ll}
{\widetilde {\omega }}^{2} = \left( \omega -m\Omega _{H} \right) ^{2} &
{\mbox {as $r_{*}\rightarrow -\infty $}}  ,
\\
\omega ^{2} & {\mbox {as  $r_{*}\rightarrow \infty $}} ,
\end{array}
\right. 
\label{eq:Kpot}
\end{equation}
where $\Omega _{H}$ is the angular speed of the event horizon.

At infinity, the potential approaches $\omega ^{2}$ as in the Schwarzschild case, and therefore the natural frequency with respect to the Schwarzschild-like time coordinate $t$ far from the black hole is simply $\omega $.
Close to the horizon, however, the potential approaches ${\widetilde {\omega }}^{2}$ and we therefore have a frequency shift, with the natural frequency of field modes close to the horizon being ${\widetilde {\omega }}$ rather than $\omega $.
This can be seen in the form of the ``in'' and ``up'' modes in the asymptotic regions:
\begin{subequations}
\label{eq:Kmodes}
\begin{align}
 R_{\omega \ell m}^{\rm {in}} (r) & \sim \left\{
 \begin{array}{ll}
 B_{\omega \ell m}^{{\mathrm {in}}} \, e^{-i{\widetilde {\omega }} r_{*}}
&  r_{*}\rightarrow -\infty ,
\\
e^{-i\omega  r_{*}}+
A_{\omega \ell m}^{{\mathrm {in}}} \,
e^{i\omega  r_{*}}
&  r_{*} \rightarrow \infty ,
 \end{array}
 \right.
 \\
R_{\omega \ell m}^{{\mathrm {up}}}(r)  & \sim 
\left\{
\begin{array}{ll}
e^{i {\widetilde {\omega }}r_{*}} 
+ A_{\omega \ell m}^{{\mathrm {up}}} \, e^{-i{\widetilde {\omega }} r_{*}} 
& r_{*}\rightarrow -\infty ,
\\
B_{\omega \ell m}^{{\mathrm {up}}} \, e^{i\omega r_{*}}
& r_{*} \rightarrow \infty ,
\end{array}
\right.
\end{align}
\end{subequations}
where $A_{\omega \ell m}^{{\mathrm {in/up}}} $ and $B_{\omega \ell m}^{{\mathrm {in/up}}}$ are complex constants.
This is our first indication that the canonical quantization of a neutral scalar field on a Kerr black hole is more complicated than on Schwarzschild space-time.
In particular, the ``in'' and ``up'' modes have different natural frequencies.
Since the ``in'' modes originate at past null infinity, the natural frequency for these modes is $\omega $, the frequency far from the black hole.
In contrast, since the ``up'' modes originate close to the past event horizon, the natural frequency for these modes is ${\widetilde {\omega }}$. 

There is a further complication. 
The Wronskian of any two linearly independent solutions of the radial equation (\ref{eq:radKerr}) is a constant, and this can be used to derive the following relation between the complex constants appearing in the asymptotic form of the ``in'' modes:
\begin{equation}
\omega \left[ 1- \left| A^{\rm {in}}_{\omega \ell m}\right| ^{2} \right] = {\widetilde {\omega }} \left| B_{\omega \ell m}^{\rm {in}} \right| ^{2} .
\end{equation}
From this, it is clear that 
\begin{equation}
\left| A^{\rm {in}}_{\omega \ell m}\right|^{2}>1 {\mbox { if }} \omega {\widetilde {\omega }} <0 .
\end{equation}
Therefore, an ``in'' mode with $\omega {\widetilde {\omega }}<0$ has a reflected part at future null infinity $A_{\omega \ell m}^{{\mathrm {in}}} \,
e^{i\omega  r_{*}}$ which has greater amplitude than the incident part $e^{-i\omega  r_{*}}$.
This is the phenomenon of {\em {superradiance}} \cite{Brito:2015oca}, and results from the wave extracting rotational energy from the black hole. 
Similarly, there are superradiant ``up'' modes with  $\left| A^{\rm {up}}_{\omega \ell m}\right|^{2}>1$ if $\omega {\widetilde {\omega }} <0$.

\subsection{Unruh state on Kerr}
\label{sec:KerrU}
To see the effect of superradiance on quantum states for the neutral scalar field, one can proceed with canonical quantization as outlined in Section~\ref{sec:intro}.
The Unruh state is constructed by taking ``in'' modes to have positive frequency with respect to Schwarzschild-like time $t$ far from the black hole (that is, $\omega >0$), while the ``up'' modes have positive frequency with respect to Kruskal time $T$ near the event horizon.
The resulting unrenormalized expectation values are \cite{Ottewill:2000qh}
\begin{equation}
\langle {\mathrm {U}} | {\hat {O}} | {\mathrm {U}} \rangle
 = 
\sum _{\ell = 0}^{\infty } \sum _{m=-\ell }^{\ell } 
\left[ 
\int _{0}^{\infty } d\omega  \,
o _{\omega  \ell m}^{{\mathrm {in}}}  +
\int _{0}^{\infty } d{\widetilde {\omega }} \, o _{\omega \ell m}^{{\mathrm {up}}} 
\coth \left( \frac {{\widetilde{\omega }}}{2T_{H}} \right) \right]  .
\end{equation}
Note that the integral over the frequency of the ``in'' modes corresponds to positive frequency for $\omega >0$ while the integral over the frequency of the ``up'' modes is positive frequency for ${\widetilde {\omega }}>0$, so that the two sets of modes are integrated over their natural frequencies.
It can also be seen that the thermal factor for the ``up'' modes involves their natural frequency ${\widetilde {\omega }}$.
As in the Schwarzschild case, the Unruh state contains an outgoing flux of thermal Hawking radiation in the ``up'' modes.

\subsection{Boulware state on Kerr}
\label{sec:KerrB}
Considering next the Boulware state, the analogue of (\ref{eq:BSchwarz}) for the neutral scalar field on Kerr space-time is
\begin{equation}
\langle {\mathrm {B}} | {\hat {O}} | {\mathrm {B}} \rangle
 = 
\sum _{\ell = 0}^{\infty } \sum _{m=-\ell }^{\ell }
\left[ 
\int _{0}^{\infty } d\omega  \,
o _{\omega  \ell m}^{{\mathrm {in}}}  +
\int _{0}^{\infty } d{\widetilde {\omega }}
\, 
o _{\omega \ell m}^{{\mathrm {up}}} 
\right] ,
\end{equation}
which results from defining positive frequency to be $\omega >0$ for the ``in'' modes and ${\widetilde {\omega }}>0$ for the ``up'' modes.
At first glance, it appears that this state is devoid of particles in the ``in'' and ``up'' modes. 
However, evaluation of the expectation value  $\langle {\mathrm {B}} | {\hat {T}}_{t}^{r} | {\mathrm {B}} \rangle $ \cite{Unruh:1974bw} reveals that the state $| {\mathrm {B}} \rangle $ in fact contains an outgoing flux of particles at future null infinity, arising from the superradiant modes with $\omega {\widetilde {\omega }}<0$.
This is the Unruh-Starobinskii effect \cite{Unruh:1974bw,Starobinsky:1973aij}, the quantum analogue of classical superradiance.

\subsection{Hartle-Hawking state on Kerr}
\label{sec:KerrH}
Finally in this section, we consider the question of the analogue of the Hartle-Hawking state $|{\mathrm {H}}\rangle $ on Kerr?  
For a neutral scalar field, the Kay-Wald theorem \cite{Kay:1988mu,Kay:1992gr}  states that there is no thermal equilibrium state on Kerr which is regular everywhere on and outside the event horizon and respects the underlying symmetries of the space-time.
In other words, there is no direct analogue of the Hartle-Hawking state for a neutral quantum scalar field on Kerr space-time. 
A natural question is whether there are nonetheless states on Kerr which possess some (but not all) of the properties we expect for a ``Hartle-Hawking''-like state. 

There are two such states in the literature.
The first \cite{Candelas:1981zv} (which we call the $|{\mathrm {CCH}}\rangle $ state), has a thermal distribution of particles in both the ``in'' and the ``up'' modes, with the natural frequencies of each set of modes appearing in the thermal factors, giving the unrenormalized expectation values \cite{Candelas:1981zv}:
\begin{equation}
\langle {\mathrm {CCH}} | {\hat {O}} | {\mathrm {CCH}} \rangle
 = 
\sum _{\ell = 0}^{\infty } \sum _{m=-\ell }^{\ell }
\left[ 
\int _{0}^{\infty } d\omega \, o _{\omega  \ell m}^{{\mathrm {in}}} 
\coth \left( \frac {\omega }{2T_{H}} \right) 
 +
\int _{0}^{\infty } d{\widetilde {\omega }} \, o _{\omega \ell m}^{{\mathrm {up}}} 
\coth \left( \frac {{\widetilde{\omega }}}{2T_{H}} \right)
\right] .
\end{equation}
This state is regular everywhere outside the event horizon, but does not represent an equilibrium state \cite{Ottewill:2000qh}. 
This may be understood from the fact that the ``in'' and ``up'' modes have different thermal factors.

An alternative state \cite{Frolov:1989jh} (which we call the $|{\mathrm {FT}}\rangle $ state) may be defined, in which  the ``in'' and ``up'' modes have the same thermal factors. 
The resulting unrenormalized expectation values are \cite{Frolov:1989jh}
\begin{equation}
\langle {\mathrm {FT}} | {\hat {O}} | {\mathrm{FT}} \rangle = 
\sum _{\ell = 0}^{\infty } \sum _{m=-\ell }^{\ell }
\left[ 
\int _{0}^{\infty } d\omega  \, o _{\omega \ell m}^{{\mathrm {in}}} 
\coth \left( \frac {{\widetilde {\omega }}}{2T_{H}} \right)
+
\int _{0}^{\infty } d{\widetilde {\omega }} \, o _{\omega \ell m}^{{\mathrm {up}}} 
\coth \left( \frac {{\widetilde {\omega }}}{2T_{H}} \right) 
\right] .
\end{equation}
This state does potentially define an equilibrium state.
However, closer inspection \cite{Ottewill:2000qh} reveals that in fact this state is divergent everywhere except on the axis of rotation of the black hole.

\subsection{Summary}
\label{sec:KerrSummary}
This brief outline has revealed some subtleties in the definition of quantum states for a neutral scalar field on a Kerr black hole background. 
In particular, there is a frequency shift in the field modes at the horizon relative to infinity, which results in the  ``in'' and ``up'' modes having different natural frequencies and the  classical phenomenon of superradiance.
The construction of the Unruh state carries over from the Schwarzschild case without difficulty, since this treats the ``in'' and ``up'' modes differently.
However, there are complications for both the Boulware and Hartle-Hawking states.
The Kerr equivalent of the Boulware state is no longer empty at future null infinity, but contains an outgoing flux of particles in the superradiant modes.
The Hartle-Hawking state is more problematic.
There is no state which has all the properties we would wish a Hartle-Hawking state to have \cite{Kay:1988mu,Kay:1992gr}. 
In the literature there are two candidate ``Hartle-Hawking''-like states \cite{Candelas:1981zv,Frolov:1989jh}.
The first \cite{Candelas:1981zv} has attractive regularity properties but is not an equilibrium state, while the second \cite{Frolov:1989jh} is potentially an equilibrium state but fails to be regular almost everywhere outside the event horizon.

\section{Charged scalar field on Reissner-Nordstr\"om black hole}
\label{sec:RN}

We now turn to the main focus of this report, the quantization of a charged scalar field on a Reissner-Nordstr\"om (RN) black hole.
This section constitutes an outline of the results in \cite{Balakumar:2020gli,Balakumar:2022yvx} to which the reader is referred for further details. 

\subsection{Charged scalar field on RN}
\label{sec:charge}
We consider the RN metric
\begin{equation}
ds^{2} = -\left(  1 - \frac{2M}{r} + \frac{Q^{2}}{r^{2}}  \right) \, dt^{2} + \left(1 - \frac{2M}{r} + \frac{Q^{2}}{r^{2}} \right) ^{-1} dr^{2}+ r^{2} \, d\theta ^{2} + r^{2} \sin ^{2} \theta \, d\varphi ^{2} ,
\end{equation}
where $M$ is the mass and $Q$ the charge of the black hole. We study a massless charged scalar field satisfying the field equation
\begin{equation}
D_{\mu} D^{\mu } \Phi =0,
\end{equation}
where $D_{\mu } = \nabla _{\mu } - iqA_{\mu }$  is the covariant derivative, $q$ the scalar field charge and $A_{\mu }$ the electromagnetic potential, whose only nonzero component is
\begin{equation}
A_{0} = -\frac{Q}{r} . 
\end{equation}
The mode solutions of the charged scalar field equation again take the form (\ref{eq:Smodes}), with the radial function $R_{\omega \ell}(r)$ satisfying a Schr\"odinger-like equation (\ref{eq:Sradial}), although the ``tortoise'' coordinate $r_{*}$ is now given by 
\begin{equation}
\frac {dr_{*}}{dr}  =   \frac {1}{1-\frac{2M}{r}+\frac{Q^{2}}{r^{2}}} .
\end{equation}
The potential $V_{\omega \ell }(r)$ is not the same as in the Schwarzschild case, and depends on the charges of the black hole and scalar field. 
In the asymptotic regions $r_{*}\rightarrow \pm \infty $ its behaviour is
\begin{equation}
V_{\omega \ell }(r)=\left\{
\begin{array}{ll}
{\widetilde {\omega }}^{2} = \left( \omega -\frac{qQ}{r_{+}} \right) ^{2} &
{\mbox {as $r_{*}\rightarrow -\infty $}} ,
\\
\omega ^{2} & {\mbox {as  $r_{*}\rightarrow \infty $}},
\end{array}
\right.   
\label{eq:RNpot}
\end{equation}  
where $r_{+}$ is the event horizon radius.
We see that, as in the Kerr case, there is a frequency shift at the event horizon relative to infinity. 

The ``in'' and ``up'' modes take the form (\ref{eq:Kmodes}) in the asymptotic regions, but with ${\widetilde {\omega }}$ defined by (\ref{eq:RNpot}) rather than (\ref{eq:Kpot}).
For modes with $\omega {\widetilde {\omega }}<0$ there is charge superradiance \cite{Bekenstein:1973mi}, with classical waves being amplified upon scattering by the black hole.
In this situation the scattered wave is extracting electrostatic rather than rotational energy from the black hole. 

Following the discussion in the previous section, we now seek to understand the effect of charge superradiance on the construction of the standard quantum states for a charged scalar field on an RN black hole, again considering unrenormalized expectation values. 
The observables of interest are the scalar condensate ${\widehat {SC}}$, current ${\hat {J}}^{\mu }$ and stress-energy tensor ${\hat {T}}_{\mu \nu }$, given in terms of the scalar field operator ${\hat {\Phi }}$ as follows:
\begin{subequations}
 \begin{align}
{\widehat {SC}} &  = \frac{1}{2} \left[ {\hat {\Phi }}{\hat {\Phi }}^{\dagger} + {\hat {\Phi }}^{\dagger}{\hat {\Phi }}\right]  ,
  \\
{\hat {J}}^{\mu } & = \frac{iq}{16\pi }\left[ {\hat {\Phi }} ^{\dagger} \left( D^{\mu } {\hat {\Phi }} \right) + \left( D^{\mu } {\hat {\Phi }} \right){\hat {\Phi}}^{\dagger} 
- {\hat {\Phi }}\left( D^{\mu } {\hat {\Phi }} \right) ^{\dagger } 
- \left( D^{\mu} {\hat {\Phi  }} \right) ^{\dagger }{\hat {\Phi }}\right]  ,
 \\ 
 {\hat {T}}_{\mu \nu } & = \frac{1}{4} \left\{  \left( D_{\mu }{\hat {\Phi }} \right)^{\dagger} D_{\nu }{\hat {\Phi }} 
	+ D_{\nu }{\hat {\Phi }} \left( D_{\mu }{\hat {\Phi }} \right)^{\dagger}
	+\left( D_{\nu }{\hat {\Phi }} \right)^{\dagger} D_{\mu }{\hat {\Phi }} 
	+ D_{\mu }{\hat {\Phi }} \left( D_{\nu }{\hat {\Phi }} \right)^{\dagger}
\right. \nonumber  \\ & ~ \left.
 \qquad -	\frac{1}{2}g_{\mu \nu }g^{\rho \sigma} \left[  \left(D_{\rho }{\hat {\Phi }}\right)^{\dagger}  D_{\sigma } {\hat {\Phi }}  + 
D_{\sigma } {\hat {\Phi }} \left(D_{\rho }{\hat {\Phi }}\right)^{\dagger} 
+\left(D_{\sigma }{\hat {\Phi }}\right)^{\dagger}  D_{\rho } {\hat {\Phi }}  + 
D_{\rho } {\hat {\Phi }} \left(D_{\sigma }{\hat {\Phi }}\right)^{\dagger}
\right]
	 \right\} .
\end{align}
\end{subequations}
We particularly focus on the fluxes of charge ${\mathcal {K}}$ and energy ${\mathcal {L}}$, which are given by the following expectation values of components of the current and stress-energy tensor:
\begin{equation}
\langle {\hat {{{J}}}}^{r} \rangle = -\frac{{\mathcal{K}}}{r^{2}}  ,
\qquad
\langle {\hat {{{T}}}}^{r}_{t} \rangle  = -\frac{{\mathcal {L}}}{r^{2}} +\frac{4\pi Q {\mathcal {K}}}{r^{3}} .
\end{equation}

\subsection{Unruh state on RN}
\label{sec:RNU}
We begin our study of quantum states for a charged scalar field on an RN black hole with an uncontroversial state, namely the Unruh state. 
The construction of this state follows that in the Kerr case, resulting in the unrenormalized expectation values \cite{Balakumar:2022yvx, Gibbons:1975kk}
\begin{equation}
\langle {\rm {U}} | {\hat {{O}}} | {\rm {U}} \rangle = 
\frac{1}{2}\sum _{\ell =0}^{\infty }\sum _{m=-\ell}^{\ell }\left[
\int _{-\infty }^{\infty }d\omega \, o _{\omega \ell m}^{\rm {in}}  
+ \int _{-\infty }^{\infty }d{\widetilde{\omega }} \,
o _{\omega \ell m}^{\rm {up}} 
\coth \left| \frac{{\widetilde{\omega }}}{2T_{H}} \right|   \,
\right] .
\end{equation}
The mode sums involve contributions from both positive and negative frequency modes, due to the charge of the scalar field. 
As in the Kerr case, there are no particles in the ``in'' modes and a thermal distribution of particles in the ``up'' modes, with the frequency ${\widetilde {\omega }}$ (the natural frequency for the ``up'' modes) in the thermal factor.
The thermal nature is clear from the expressions for the fluxes of charge and energy in this state \cite{Balakumar:2022yvx,Gibbons:1975kk}:
\begin{subequations}
\label{eq:Ufluxes}
\begin{align}
{\mathcal {K}}_{{\mathrm {U}}} &  =  \frac{q}{64 \pi^{3}} \sum_{\ell = 0}^\ell  \int_0^\infty d \omega 
\left(2 \ell + 1 \right)\omega  
\left[ \frac{ {\left| {B}^\mathrm{up}_{\omega \ell} \right|}^2}{{\widetilde {\omega }}\left( \exp{ \left[ \frac{ {{\widetilde {\omega }}}}{T_{H}} \right]} - 1 \right) }
- \frac{ {\left| {B}^\mathrm{up}_{-\omega \ell} \right|}^2}{{\overline {\omega }}\left( \exp{ \left[ \frac{ {\overline {\omega }}}{T_{H} }  \right]} - 1 \right)  } \right]   ,
\\
{\mathcal {L}}_{{\mathrm {U}}} &  =  \frac{1}{16 \pi^{2}} \sum_{\ell = 0}^\ell  \int_0^\infty d \omega 
\left(2 \ell + 1 \right)\omega  ^{2} 
\left[ \frac{ {\left| {B}^\mathrm{up}_{\omega \ell} \right|}^2}{{\widetilde {\omega }}\left( 
 \exp\left[ \frac{ {\widetilde {\omega }}}{T_{H} } \right] - 1 \right) }
	+ \frac{ {\left|{B}^\mathrm{up}_{-\omega \ell} \right|}^2}{{\overline {\omega }}\left( \exp \left[ \frac{ {\overline {\omega }}}{T_{H}}  \right] - 1 \right)}  \right]    ,
\end{align}
\end{subequations}
where for negative frequency modes the term in the thermal factor is
\begin{equation}
    {\overline {\omega }} = \omega + \frac{qQ}{r_{+}} .
    \label{eq:omegabar}
\end{equation}
The fluxes (\ref{eq:Ufluxes}) contain contributions from both positive and negative frequency modes. For positive frequency modes, there is thermal emission with an effective chemical potential $qQ/r_{+}$ \cite{Hawking:1974sw,Gibbons:1975kk}. The sign of the effective chemical potential is reversed for negative frequency modes. 

\subsection{Boulware state on RN}
\label{sec:RNB}
We now turn to the construction of a ``Boulware''-like state for a charged scalar field on RN space-time.
We quantize the ``in'' and ``up'' modes with respect to their natural frequencies, defining positive frequency to be $\omega >0$ for the ``in'' modes and ${\widetilde {\omega }}>0$ for the ``up'' modes. 
We denote the resulting vacuum state $|{\rm {B}}_{1}\rangle $. 
The construction gives the following unrenormalized expectation values \cite{Balakumar:2022yvx}
\begin{equation}
\langle {\rm {B}}_{1} |{\hat {{O}}} | {\rm {B}}_{1} \rangle = 
\frac{1}{2}\sum _{\ell =0}^{\infty }\sum _{m=-\ell}^{\ell }\left[
\int _{-\infty }^{\infty }d\omega \, o _{\omega \ell m}^{{\rm {in}}}  
+ \int _{-\infty }^{\infty }d{\widetilde{\omega }} \,
o _{\omega \ell m}^{{\rm {up}}}
\right] .
\label{eq:B1RN}
\end{equation}
As in the Kerr case, this state is not empty at future null infinity.
There is an outgoing flux of charge and energy in the superradiant ``up'' modes
\cite{Balakumar:2020gli}:
\begin{subequations}
\begin{align}
{\mathcal {K}}_{{\mathrm {B}}_{1}} & =  
\frac{q}{64\pi ^{3}}\sum _{\ell =0}^{\infty } 
\int _{\min \{\frac{qQ}{r_{+}},0\}}^{\max \{ \frac{qQ}{r^{+}},0\}} d\omega 
\frac{\omega }{\left| {\widetilde {\omega }} \right| }\left( 2\ell + 1 \right) \left| B^{\rm {up}}_{\omega \ell } \right| ^{2}  ,
 \\
{\mathcal {L}}_{{\mathrm {B}}_{1}} & =    \frac{1}{16 \pi ^{2}} \sum _{\ell =0}^{\infty } 
\int _{\min \{\frac{qQ}{r_{+}},0\}}^{\max \{ \frac{qQ}{r^{+}},0\}} d\omega 
\frac{\omega ^{2}}{\left| {\widetilde {\omega }} \right| }\left( 2\ell + 1 \right) \left| B^{\rm {up}}_{\omega \ell } \right| ^{2} .
\end{align}
\end{subequations}
The question is then whether it is possible to construct a state which is as empty as possible at both past and future null infinity.
Such a state, denoted $|{\rm {B}}\rangle $, is constructed in \cite{Balakumar:2022yvx}, using creation and annihilation operators which satisfy nonstandard commutation relations (see \cite{Balakumar:2022yvx} for details). 
As far as unrenormalized expectation values are concerned, the superradiant ``up'' modes make a contribution which has the opposite sign to that in (\ref{eq:B1RN}) \cite{Balakumar:2022yvx}:
\begin{equation}
\langle {\rm {B}} | {\hat {{O}}} |  {\rm {B}}\rangle =  ~
\frac{1}{2} \sum _{\ell =0}^{\infty }\sum _{m=-\ell}^{\ell }
\left[ 
\int _{-\infty }^{\infty } d\omega  \left[ 
o _{\omega \ell m}^{\rm {in}}  +
o _{\omega \ell m}^{\rm {up}} 
\right] 
- 2\int _{\min \{ 0, \frac{qQ}{r_{+}}\} } ^{\max \{ 0, \frac{qQ}{r_{+}} \} }
d\omega  \,
o _{\omega \ell m}^{\rm {up}}     \right]  .
\end{equation} 
The fluxes of charge and energy in this state vanish:
\begin{equation}
{\mathcal {K}}_{\mathrm {B}} =0,   \qquad 
{\mathcal {L}}_{\mathrm {B}}  =0,
\end{equation}
and therefore the state $|{\rm {B}}\rangle $ is as empty as possible at both past and future null infinity. 

As with the Boulware state on Schwarzschild space-time, the state $|{\rm {B}}\rangle $ diverges on the event horizon of the RN black hole. 
It is argued in \cite{Ottewill:2000qh} that there is no vacuum state for a neutral quantum scalar field on Kerr space-time which is as empty as possible at both past and future null infinity. 
This appears to also be the case for a charged quantum scalar field on RN space-time, since the ``Boulware''-like state $|{\rm {B}}\rangle $
is not a conventional vacuum state, as its construction relies on nonstandard commutation relations \cite{Balakumar:2022yvx}.

\subsection{Hartle-Hawking state on RN}
\label{sec:RNH}
Finally we consider the construction of an analogue of the Hartle-Hawking state for a charged quantum scalar field on an RN black hole.
Following our discussion of ``Hartle-Hawking''-like states on Kerr space-time, we begin with a state $|{\mathrm {H}}_{1}\rangle $, which is the analogue of the $|{\mathrm {CCH}}\rangle $ state on Kerr. 
Unrenormalized expectation values in this state take the form \cite{Balakumar:2022yvx}
\begin{equation}
\langle {\mathrm {H}}_{1} | {\hat {O}} | {\mathrm {H}}_{1}  \rangle = 
\sum _{\ell = 0}^{\infty } \sum _{m=-\ell }^{\ell }
\left[
\int _{-\infty }^{\infty } d\omega  \,
o _{\omega  \ell m}^{{\mathrm {in}}} 
\coth \left| \frac {\omega }{2T_{H}} \right| 
+
\int _{-\infty  }^{\infty } d{\widetilde {\omega }} \,
o _{\omega \ell m}^{{\mathrm {up}}} 
\coth \left| \frac {{\widetilde{\omega }}}{2T_{H}} \right|  \,
\right] .
\end{equation}
Both the ``in'' and ``up'' modes contain a thermal distribution of particles at the Hawking temperature $T_{H}$, however the frequencies in the thermal factors are different. 
The ``in'' modes are thermalized with respect to their natural frequency $\omega $, while the ``up'' modes are thermalized with respect to their natural frequency ${\widetilde {\omega }}$.  

As might be expected, the state $|{\mathrm {H}}_{1}\rangle $ is not an equilibrium state. 
The fluxes of charge and energy in this state are \cite{Balakumar:2022yvx}:
\begin{subequations}
\begin{align}
{\mathcal {K}}_{{\mathrm {H}}_{1}} & = {\mathcal {K}}_{{\mathrm {U}}}
-\frac{q}{64 \pi^{3}} \sum_{\ell = 0}^\infty  \int_0^\infty d \omega \,
\frac{\left(2 \ell + 1 \right){\widetilde {\omega }}^{2}}{\omega \! \left( \exp \left[ \frac{ \omega }{T_{H} }  \right] -1\right) }
\left[
\frac{1}{{\widetilde {\omega }}}\left| B_{\omega \ell }^{\rm {in}}\right| ^{2} 
- \frac {1}{{\overline {\omega }}}\left| B_{-\omega \ell }^{\rm {in}}\right| ^{2}
\right]  ,
\\
{\mathcal {L}}_{{\mathrm {H}}_{1}} & = {\mathcal {L}}_{{\mathrm {U}}}
-\frac{1}{16 \pi^{2}} \sum_{\ell = 0}^\infty  \int_0^\infty d \omega \,
\frac{\left(2 \ell + 1 \right){\widetilde {\omega }}^{2}}{\left( \exp \left[ \frac{\omega }{T_{H}}  \right] -1\right) }
\left[
\frac{1}{{\widetilde {\omega }}}\left| B_{\omega \ell }^{\rm {in}}\right| ^{2} 
+ \frac {1}{{\overline {\omega }}}\left| B_{-\omega \ell }^{\rm {in}}\right| ^{2}
\right]  ,
\end{align}
\end{subequations}
where ${\mathcal {K}}_{{\mathrm {U}}}$ and ${\mathcal {L}}_{\mathrm {U}}$ are, respectively, the fluxes of charge and energy in the Unruh state (\ref{eq:Ufluxes}).
In the ``up'' modes, the fluxes are the same as those for the Unruh state $|{\mathrm {U}}\rangle $, since the thermalization of the ``up'' modes is the same in these two states.
The state $|{\mathrm{H}}_{1}\rangle $ has additional thermal fluxes for the ``in'' modes, with the thermal factor containing the natural frequency $\omega $ for these modes.

Again following the constructions in the Kerr case, our second ``Hartle-Hawking''-like state, $|{\mathrm {H}}_{2}\rangle $,  has identical thermal factors for both the ``in'' and ``up'' modes, with 
unrenormalized expectation values given by \cite{Balakumar:2022yvx}
\begin{equation}
\langle {\mathrm {H}}_{2} | {\hat {O}} | {\mathrm {H}}_{2} \rangle
= 
\sum _{\ell = 0}^{\infty } \sum _{m=-\ell }^{\ell }
\left[ 
\int _{-\infty }^{\infty } d{\widetilde {\omega  }} \, 
o _{\omega  \ell m}^{{\mathrm {in}}} 
\coth \left| \frac {{\widetilde {\omega }}}{2T_{H}} \right| 
 +
\int _{-\infty  }^{\infty } d{\widetilde {\omega }} \, 
o _{\omega \ell m}^{{\mathrm {up}}}  
\coth \left| \frac {{\widetilde{\omega }}}{2T_{H}} \right| \,
\right] .
\end{equation}
The state $|{\mathrm {H}}_{2}\rangle $ is the analogue, for a charged quantum scalar field on an RN black hole, of the state $|{\mathrm {FT}}\rangle $ for a neutral quantum scalar field on a Kerr black hole.

Unlike the state $|{\mathrm {FT}}\rangle $ on Kerr, the state $|{\mathrm {H}}_{2}\rangle $ is not an equilibrium state.
It has nonzero fluxes of charge and energy in the superradiant ``in'' modes:
\begin{subequations}
\begin{align}
{\mathcal {K}}_{{\mathrm {H}}_{2}} & = 
\frac{q}{64\pi ^{3}} \sum _{\ell =0}^{\infty }
\int _{\min \left\{\frac{qQ}{r_{+}},0\right\} } ^{\max\left\{\frac{qQ}{r_{+}},0\right\} } d\omega  \,
\frac{\left| {\widetilde {\omega }} \right| }{\omega } \left( 2\ell + 1 \right)
\left| {{B}}_{\omega \ell }^{\mathrm {in}} \right| ^{2} 
\coth \left| \frac{ {\widetilde {\omega }}}{2T_{H}}\right| ,
\\
{\mathcal {L}}_{{\mathrm {H}}_{2}} & = 
\frac{1}{16\pi ^{2}} \sum _{\ell =0}^{\infty }
\int _{\min \left\{\frac{qQ}{r_{+}},0\right\} } ^{\max\left\{\frac{qQ}{r_{+}},0\right\} } d\omega  \,
\left| {\widetilde {\omega }} \right|  \left( 2\ell + 1 \right) 
\left| {{B}}_{\omega \ell }^{\mathrm {in}} \right| ^{2} 
\coth \left| \frac{ {\widetilde {\omega }}}{2T_{H} }\right|  .
\end{align}
\end{subequations}
In \cite{Balakumar:2022yvx}, the properties of the state $|{\mathrm {H}}_{2}\rangle $ are studied by considering the differences in expectation values between this state and the uncontroversial Unruh state $|{\mathrm {U}}\rangle $. 
The differences in the expectation values of the current and stress-energy tensor are regular everywhere outside the event horizon.
Since it is anticipated that the Unruh state $|{\mathrm {U}}\rangle $ is also regular everywhere outside the event horizon, this may lead us to deduce that the state $|{\mathrm {H}}_{2}\rangle $ is also regular everywhere outside the event horizon.
However, there is a problem when we consider the difference in expectation values of the scalar condensate:
\begin{equation}
\langle {\mathrm {H}}_{2} | {\widehat {{SC}}} | {\mathrm {H}}_{2} \rangle  
- \langle {\mathrm {U}} | {\widehat {{SC}}} | {\mathrm {U}} \rangle
=
\sum _{\ell =0}^{\infty }\sum _{m=-\ell}^{\ell }
\int _{-\infty }^{\infty }d\omega  \,
\frac{1}{\exp \left| \frac{ {\widetilde {\omega }}}{T_{H} } \right| -1 } 
\left| \phi ^{\rm {in}}_{\omega \ell m} \right|^{2}.
\label{eq:FTSC}
\end{equation}
The integrand in (\ref{eq:FTSC}) has a pole when ${\widetilde {\omega }}=0$, and the integral is not convergent. 
Given the expected regularity of the Unruh state $|{\mathrm {U}}\rangle $, we infer that the state $|{\mathrm {H}}_{2}\rangle $ is divergent everywhere. 
This situation is similar to that for the state $|{\mathrm {FT}}\rangle $ on Kerr, whose divergent nature is revealed by considering the vacuum polarization (the scalar condensate for a neutral scalar field) \cite{Ottewill:2000qh}.

Thus far we have postulated two ``Hartle-Hawking''-like states: the first, $|{\mathrm {H}}_{1}\rangle $, is regular everywhere outside the event horizon but is not an equilibrium state; the second, $|{\mathrm {H}}_{2}\rangle $, is also not an equilibrium state but is divergent everywhere outside the event horizon.
In \cite{Balakumar:2022yvx}, a third potential ``Hartle-Hawking''-like state, $|{\mathrm {H}}\rangle $, is constructed.
Like the state $|{\mathrm {B}}\rangle $, the construction of the state $|{\mathrm {H}}\rangle $ relies on the use of creation and annihilation operators satisfying nonstandard commutation relations. 
This has the effect of reversing the sign of the contribution of the superradiant ``in'' modes to unrenormalized expectation values, compared to those in the state $|{\mathrm {H}}_{2}\rangle $, yielding:
\begin{multline}
\langle {\rm {H}} |{\hat {{O}}}|  {\rm {H}} \rangle = 
\frac{1}{2} \sum _{\ell =0}^{\infty }\sum _{m=-\ell}^{\ell }
\left[  \int _{-\infty }^{\infty }d{\widetilde{\omega }}  \,
\left[
o _{\omega \ell m}^{\rm {in}} +
o _{\omega \ell m}^{\rm {up}} \right]
\coth \left| \frac{ {\widetilde{\omega }}}{2T_{H} } \right|
\right. 
\\ \left. 
- 2 \int _{\min \{ 0, \frac{qQ}{r_{+}}\} } ^{\max \{ 0, \frac{qQ}{r_{+}} \} }
d\omega  \,
o _{\omega \ell m}^{\rm {in}}  \coth \left| \frac{ {\widetilde{\omega }}}{2T_{H}} \right|   \,
\right] .
\end{multline}
Unlike the states $|{\mathrm {H}}_{1}\rangle $ and $|{\mathrm {H}}_{2}\rangle $, our new state $|{\mathrm {H}}\rangle $ is an equilibrium state, with vanishing fluxes of both charge and energy \cite{Balakumar:2022yvx}:
\begin{equation}
{\mathcal {K}}_{\mathrm {H}} =0,  \qquad {\mathcal {L}}_{\mathrm {H}} = 0.
\end{equation}
This is one of the properties we were seeking. The other is regularity everywhere outside the event horizon.

Based on our experience with the states $|{\mathrm {H}}_{1}\rangle $ and $|{\mathrm {H}}_{2}\rangle $, the key quantity to examine is the scalar condensate. 
We find the following expression for the difference in expectation values of the scalar condensate in the state $|{\mathrm {H}}\rangle $ and the Unruh state:
\begin{equation}
\langle {\rm {H}} | {\widehat {{SC}}} | {\rm {H}} \rangle  
- \langle {\rm {U}}  | {\widehat {{SC}}} | {\rm {U}} \rangle
= 
\sum _{\ell =0}^{\infty } \sum _{m=-\ell }^{\ell }
\int _{0}^{\infty } d\omega  
\left[ 
\frac{\left| \phi _{\omega \ell m}^{\rm {in}}\right|^{2}}{\exp \left(  \frac{{\widetilde {\omega }}}{T_{H} } \right) -1}
+ \frac{\left| \phi _{-\omega \ell m}^{\rm {in}}\right|^{2}}{\exp \left(  \frac{ {\overline {\omega }}}{T_{H}} \right) -1} 
\right] ,
\label{eq:HSC}
\end{equation}
where ${\widetilde {\omega }}$ is given by (\ref{eq:RNpot}) and ${\overline {\omega }}$ by (\ref{eq:omegabar}).
This clearly has a thermal distribution of particles in the ``in'' modes, with the chemical potential changing signs for the negative frequency modes relative to the positive frequency modes (compare (\ref{eq:Ufluxes}) for the fluxes in the ``up'' modes in the Unruh state).
Each integral in (\ref{eq:HSC}) has a pole in the denominator when either ${\widetilde {\omega }}=0$ or ${\overline {\omega }}=0$. 
However, unlike the poles arising in the integrand in (\ref{eq:FTSC}),
the Cauchy principal values of the integrals in (\ref{eq:HSC}) exist, leading to a regular scalar condensate expectation value in the $|{\mathrm {H}}\rangle $ state \cite{Balakumar:2022yvx}. 
Further numerical investigations \cite{Balakumar:2022yvx} reveal that the differences in expectation values for the scalar condensate, current and stress-energy tensor between the state $|{\mathrm {H}} \rangle $ and the Unruh state are all regular everywhere outside the event horizon.
We therefore deduce that the state $|{\mathrm {H}} \rangle $ is, like the Unruh state, regular everywhere outside the event horizon.

\section{Conclusions}
\label{sec:conc}

The standard Unruh \cite{Unruh:1976db}, Boulware \cite{Boulware:1974dm} and Hartle-Hawking \cite{Hartle:1976tp} states for quantum fields on a Schwarzschild black hole were constructed in the 1970s and since then have been extensively studied. 
Extending these definitions to more general black holes has proven to be nontrivial.
On a rotating Kerr black hole, for a neutral scalar field there is no Hartle-Hawking state \cite{Kay:1988mu, Kay:1992gr}, and attempts to define a state with some (but not all) of the properties of a Hartle-Hawking state result in either a state \cite{Candelas:1981zv} which is regular outside the event horizon but not an equilibrium state, or an equilibrium state \cite{Frolov:1989jh} which is divergent almost everywhere \cite{Ottewill:2000qh}. 
A key property of bosonic quantum fields on Kerr space-time is the classical phenomenon of superradiance \cite{Brito:2015oca}, by which an incident wave is amplified upon scattering by the black hole.
In this note we have described superradiance as resulting from a frequency shift in scalar field modes at the event horizon relative to infinity. 
This frequency shift complicates the construction of quantum states via canonical quantization, since the first step is to split the field modes into ``positive'' and ``negative'' frequencies. 

A frequency shift, and consequent superradiance, also occurs for a charged scalar field on a Reissner-Nordstr\"om black hole. 
In analogy with the Kerr case, we have found that the definition of the standard quantum states in this set-up is also complicated by the superradiant modes. 
As well as analogues of the ``Hartle-Hawking''-like states \cite{Candelas:1981zv,Frolov:1989jh} discussed above for a neutral scalar field on Kerr, we have postulated two new states, $|{\mathrm {B}}\rangle $ and $|{\mathrm {H}}\rangle$.
The state $|{\mathrm {B}}\rangle $ is as empty as possible at both past and future null infinity, and is analogous to the Boulware state on Schwarzschild.
The state $|{\mathrm {H}}\rangle$ is the analogue of the Hartle-Hawking state on Schwarzschild --- it is a thermal equilibrium state and appears to be regular everywhere outside the event horizon.
However, we anticipate that a Kay-Wald-like theorem \cite{Kay:1988mu,Kay:1992gr} will still hold for a charged scalar field on Reissner-Nordstr\"om space-time, since the construction of both the states $|{\mathrm {B}}\rangle $ and $|{\mathrm {H}}\rangle$ relies on using creation and annihilation operators which do not satisfy the standard commutation relations (see \cite{Balakumar:2022yvx} for details).
Therefore, it is likely that these states do not satisfy the assumptions of a Kay-Wald-like theorem \cite{Kay}.

\section*{Acknowledgments}
We thank the organizers of the workshop ``Avenues of Quantum Field Theory in Curved Space-Time'' for a very interesting and stimulating meeting, at which this work was presented. 
	We thank Lu\'is C.~B.~Crispino and Bernard Kay for invaluable discussions on this project.
	V.B.~thanks STFC for the provision of a studentship supporting this work,
	and the Faculdade de F\'isica, Universidade Federal do Par\'a, for hospitality while this work was in progress.
The
work of V.B.~is also supported by an EPSRC Mathematical Sciences
Research Associateship.
	The work of R.P.B.~is financed in part by Coordena\c{c}\~ao de Aperfei\c{c}oamento de Pessoal de N\'ivel Superior (CAPES, Brazil) - Finance Code 001 and by 
	Conselho Nacional de Desenvolvimento Cient\'ifico e Tecnol\'ogico (CNPq, Brazil).
	The work of E.W.~is supported by the Lancaster-Manchester-Sheffield Consortium for Fundamental Physics under STFC grant ST/T001038/1.
	This research has also received funding from the European Union's Horizon 2020 research and innovation program under the H2020-MSCA-RISE-2017 Grant No.~FunFiCO-777740.  

\section*{References}
\bibliographystyle{iopart-num}
\bibliography{genoa}

\providecommand{\newblock}{}
\begin{thebibliography}{10}
\expandafter\ifx\csname url\endcsname\relax
  \def\url#1{{\tt #1}}\fi
\expandafter\ifx\csname urlprefix\endcsname\relax\def\urlprefix{URL }\fi
\providecommand{\eprint}[2][]{\url{#2}}

\bibitem{Hawking:1974sw}
Hawking S~W 1975 {\em Commun.\ Math.\ Phys.\/} {\bf 43} 199--220

\bibitem{Unruh:1976db}
Unruh W~G 1976 {\em Phys.\ Rev.\ D\/} {\bf 14} 870--892

\bibitem{Boulware:1974dm}
Boulware D~G 1975 {\em Phys. Rev. D\/} {\bf 11} 1404--1423

\bibitem{Hartle:1976tp}
Hartle J~B and Hawking S~W 1976 {\em Phys. Rev. D\/} {\bf 13} 2188--2203

\bibitem{Israel:1976ur}
Israel W 1976 {\em Phys. Lett. A\/} {\bf 57} 107--110

\bibitem{Novikov:1989sz}
Novikov I~D and Frolov V~P 1989 {\em {Physics of black holes}\/} (Dordrecht,
  Netherlands: Kluwer Academic)

\bibitem{Decanini:2005eg}
Decanini Y and Folacci A 2008 {\em Phys. Rev. D\/} {\bf 78} 044025
  (\textit{Preprint} \eprint{gr-qc/0512118})

\bibitem{Balakumar:2019djw}
Balakumar V and Winstanley E 2020 {\em Class.\ Quant.\ Grav.\/} {\bf 37} 065004
  (\textit{Preprint} \eprint{1910.03666})

\bibitem{Brito:2015oca}
Brito R, Cardoso V and Pani P 2015 {\em {Superradiance}: {energy extraction,
  black-hole bombs and implications for astrophysics and particle physics}\/}
  (Springer) (\textit{Preprint} \eprint{1501.06570})

\bibitem{Ottewill:2000qh}
Ottewill A~C and Winstanley E 2000 {\em Phys. Rev. D\/} {\bf 62} 084018
  (\textit{Preprint} \eprint{gr-qc/0004022})

\bibitem{Unruh:1974bw}
Unruh W~G 1974 {\em Phys.\ Rev.\ D\/} {\bf 10} 3194--3205

\bibitem{Starobinsky:1973aij}
Starobinsky A~A 1973 {\em Sov.\ Phys.\ JETP\/} {\bf 37} 28--32

\bibitem{Kay:1988mu}
Kay B~S and Wald R~M 1991 {\em Phys.\ Rept.\/} {\bf 207} 49--136

\bibitem{Kay:1992gr}
Kay B~S 1993 {\em J.\ Math.\ Phys.\/} {\bf 34} 4519--4539

\bibitem{Candelas:1981zv}
Candelas P, Chrzanowski P and Howard K~W 1981 {\em Phys. Rev. D\/} {\bf 24}
  297--304

\bibitem{Frolov:1989jh}
Frolov V~P and Thorne K~S 1989 {\em Phys. Rev. D\/} {\bf 39} 2125--2154

\bibitem{Balakumar:2020gli}
Balakumar V, Winstanley E, Bernar R~P and Crispino L~C~B 2020 {\em Phys. Lett.
  B\/} {\bf 811} 135904 (\textit{Preprint} \eprint{2010.01630})

\bibitem{Balakumar:2022yvx}
Balakumar V, Bernar R~P and Winstanley E 2022 {\em Phys. Rev. D\/} {\bf 106}
  125013 (\textit{Preprint} \eprint{2205.14483})

\bibitem{Bekenstein:1973mi}
Bekenstein J~D 1973 {\em Phys. Rev. D\/} {\bf 7} 949--953

\bibitem{Gibbons:1975kk}
Gibbons G~W 1975 {\em Commun.\ Math.\ Phys.\/} {\bf 44} 245--264

\bibitem{Kay}
Kay B~S 2022 private communication

\end{thebibliography}

\end{document}